\documentclass[12pt]{iopart}
\usepackage{iopams,cite}
\def\bea{\begin{eqnarray}}
\def\eea{\end{eqnarray}}
\def\beq{\begin{equation}}
\def\eq{\end{equation}}

\begin{document}

\title[Lie groups and numerical solutions of differential equations]{Lie groups and numerical solutions of differential equations: Invariant discretization versus differential approximation}

\author{D. Levi$^*$ and P. Winternitz$^+$}

\address{$^*$ Dipartimento di Matematica e Fisica, Universit\'a degli Studi Roma Tre and INFN, Sezione Roma Tre, Via della Vasca Navale 84, 00184 Roma, Italy\\ 
$^+$ Centre de Recherches Math\'ematiques, Universit\'e de Montr\'eal,
C.P. 6128, succ. Centre-ville, Montr\'eal, QC, H3C 3J7, Canada}

\begin{abstract}
We briefly review two different methods of applying Lie group theory in the numerical solution of ordinary differential equations. On specific examples we show how the symmetry preserving discretization provides difference schemes for which the "first differential approximation" is invariant under the same Lie group as the original ordinary differential equation. 
 \end{abstract}


\maketitle

Dedicated to Professor Miloslav Havl\'i\v{c}ek on the occasion of his 75 birthday.

\section{Introduction}

Lie group theory provides powerful tools for solving ordinary and partial differential equations, specially nonlinear ones \cite{1,2,3}. The standard approach is to find the Lie point symmetry group $G$ of the equation and then look for invariant solutions, i.e. solutions that are invariant under some subgroup $G_0 \subset G$. For Ordinary Differential Equations (ODE's) this leads to a reduction of the order of the equation (without any loss of information). If the dimension of the symmetry group is large enough (at least equal to the order of the equation) and the group is solvable, then the order of the equation can be reduced to zero. This can be viewed as obtaining the general solution of  the equation, explicitly or implicitly. 

However, for an ODE obtaining an implicit solution is essentially equivalent to replacing the differential equation by an algebraic or a functional one.From the point of view of visualizing the solution, or presenting a graph of the solution, it maybe easier to solve the ODE numerically than to do the same for the functional equation.

For Partial Differential Equations (PDE's) symmetry reduction reduces the number of independent variables in the equation and leads to particular solutions, rather then the general one.

Both for ODE's and PDE's with nontrivial symmetry groups it may still be necessary to resort to numerical solutions. The question then arises of making good use of the group $G$. Any numerical method involves replacing the differential equation by a difference one. In standard discretizations no heed is paid to the symmetry group $G$ and some, or all of the symmetries are lost. Since the Lie point symmetry group encodes many of the properties of the solution space of a differential equation, it seems desirable to preserve it, or at least some of its features in the discretization process.

Two different methods for incorporating symmetry concepts into the discretization of differential equations exist in the literature. One was proposed and explored by Shokin and Yanenko \cite{4,5,6,7,8}  for PDE's and has been implemented in several recent studies \cite{9,10}.
It is called the {\it differential approximation method} and the basic idea is the following. A uniform orthogonal lattice in $x_i$ is introduced and the considered differential equation 
\beq \label{e1}
E(\vec x, u, u_{x_i}, u_{x_i,x_j}, \cdots )=0,
\eq
is approximated by some difference equation $E^{\Delta}=0$. The derivatives are replaced by discrete derivatives. All known and unknown functions in (\ref{e1}) are then expanded in Taylor series about some reference point ($\vec x$), in terms of the lattice spacings. In the simplest case of 2 variables, $x$ and $t$, we have $\sigma=x_{n+1}-x_n, \, \tau=t_{n+1}-t_n$
\bea \label{e2}
&&E^{\Delta}(x,t,u,\Delta_x u, \Delta_t u, \Delta_{xx} u, \Delta_{tt} u, \Delta_{xt} u, \cdots) = \\ \nonumber
&& = E + \sigma E_1 + \tau E_2 + \sigma^2 E_3 + 2 \sigma \tau E_4 + \tau^2 E_5 +\cdots.
\eea
The expansion on the right hand side of (\ref{e2}) is a "differential approximation" of the difference equation 
\bea \label{e3}
E^{\Delta} = 0.
\eea
The "zero order differential approximation" of (\ref{e3}) is the original differential equation (\ref{e1}) and hence is invariant under the symmetry group $G$. Keeping terms of order $\sigma$ or $\tau$ in (\ref{e2}) we obtain the first differential approximation, etc. The idea is to take a higher order differential approximation, at least the first order one, and require that is also invariant under $G$, or at least under a subgroup $G_0 \subset G$. This is done by constructing different possible difference schemes approximating eq. (\ref{e1}) and choosing among them the one for which the first (or higher) differential approximation has the "best" symmetry properties.

The second approach, the {\it invariant discretization} method is part of a program devoted to the study of continuous symmetries of discrete equations, i.e. the application of Lie groups to difference equations \cite{11,12,13,14,15,16,17,18,19,20,21,22,23,24}. As far as applications to the numerical solutions of differential equations are concerned, the idea, originally due to Dorodnitsyn \cite{12,13}, is to start from the differential equation, its symmetry group $G$ and the Lie algebra $L$ of $G$, realized by vector fields. The differential equation can then be expressed in terms of differential invariants of $G$. The differential equation is then approximated by a finite difference scheme that is constructed so as to be invariant under the same group $G$ as the differential equation.  The difference scheme will consist of several equations establishing relations between points in the space of independent and dependent variables. These equations determine both the evolution of the dependent variables and the form of the lattice. The equations are written in terms of group invariants of the group $G$ acting via its prolongation to all points on the lattice (rather then to derivatives of the dependent functions). As pointed out by P. Olver this amounts to prolonging the group action  to "multi--space" for difference equations \cite{24} rather then to "jet space" as for differential equations \cite{25}.

The purpose of this article is to compare the two different methods of incorporating Lie symmetries into the numerical analysis of differential equations. For simplicity we restrict ourselves to the case of ODE's  and analyze difference schemes that were used in recent articles \cite{26,27,28} to solve numerically some third order nonlinear ODE's with three or four dimensional symmetry algebras. In this article we take  invariant difference schemes (on symmetry adapted lattices) and construct its first differential approximation. We then verify that in all examples this first differential approximation is invariant under the entire symmetry group $G$. 

\section{Differential approximations of ordinary difference equations and invariant discretization of ODE's.}

Let us consider the case of a third order ODE
\beq \label{e4}
E \equiv E(x,y,y',y'',y''')=0,
\eq
(the generalization to order $n\ge 3$ is straightforward). We can approximate (\ref{e4}) on a 4 point stencil with points ($x_k, y_k$), ($k=n-1,n,n+1,n+2$). Alternative coordinates on the stencil are the coordinates of one reference point, say ($x_n,y_n$), the distances between the points, and the discrete derivatives up to order 3,
\bea \label{e5}
&&\big \{ x_n,\, y_n,\; h_{n+k}=x_{n+k}-x_{n+k-1},\\ \nonumber &&p^{(1)}_{n+1}=\frac{y_{n+1}-y_n}{x_{n+1}-x_n}, \;
 p^{(2)}_{n+2}=2 \frac{ p^{(1)}_{n+2}- p^{(1)}_{n+1}}{x_{n+2}-x_n},\\ \nonumber && p^{(3)}_{n+3}=3 \frac{ p^{(2)}_{n+3}- p^{(2)}_{n+2}}{x_{n+3}-x_n}\big  \}.
\eea
The one dimensional lattice can be chosen to be uniform and then 
\beq \label{e6}
h_{n+1}=h_n \equiv h,
\eq
or some other distribution of points can be chosen.

The Lie point symmetry group $G$ transform the variables ($x,y$) into $(\tilde x, \tilde y)=\big (\Lambda(x,y), \Omega(x,y) \big )$ and its Lie point symmetry algebra $L$ is represented by vector fields of the form
\beq \label{e7}
\hat X_{\mu} = \xi_{\mu}(x,y) \partial_x +\phi_{\mu}(x,y) \partial_y, \quad 1 \le \mu \le \mbox{dim}L.
\eq
The vector fields must be prolonged in the standard manner \cite{1} to derivatives
\bea \label{e8}
\mbox{pr} \hat X_{\mu} &=& \hat X_{\mu} + \phi^{x}(x,y,y_x)\partial_{y_x} +\\ \nonumber  &+&\phi^{xx}(x,y,y_x,y_{xx})\partial_{y_{xx}} + \cdots,
\eea
when acting on a differential equation, or to all points when acting on a difference scheme \cite{15,18,20}
\bea \label{e8a}
\mbox{pr}^{\Delta} \hat X_{\mu} &=& \sum_i \Big [ \xi_{\mu} (x_i, y_i ) \partial_{x_i} + \phi_{\mu}(x_i, y_i) \partial_{y_i} \Big ], 
\eea
In the differential approximation method we start with a difference equation, usually on a uniform lattice (\ref{e6})
\beq \label{e9}
E^{\Delta}(x_n,h_{n+1},h_{n+2},y_n,p^{(1)}_{n+1},p^{(2)}_{n+2} ,p^{(3)}_{n+3})=0,
\eq
and expand it into a Taylor series in the spacing $h$:
\beq \label{e10}
E^{\Delta} = E_0 + h E_1 + h^2 e_2 + \cdots =0.
\eq
For any difference equation approximating the differential equation $E_0=0$ the lowest order term will be invariant under the group $G$.  Different schemes (\ref{e9}) can then be compared with respect to the invariance of the first differential approximation
\beq \label{e11}
E_0 + h E_1=0,
\eq
or of some higher order differential approximations. Better results can be expected for schemes from which (\ref{e11}) is invariant under all of $G$ or under some subgroup $G_0 \subset G$ that is relevant for the problem.

When studying the invariance of (\ref{e11}) it must be remembered that the prolongations of $\hat X_{\mu}$ also act on the lattice parameters $h$.

In the invariant discretization method one constructs an invariant difference scheme
\bea \label{e12}
&&E^{\Delta}_a(x_n,y_n,h_n,h_{n+1},h_{n+2},y_n,\\ \nonumber&&p^{(1)}_{n+1},p^{(2)}_{n+2} ,p^{(3)}_{n+3})  =0, \qquad a=1,2 \\ \label{e13}
&& \mbox{pr}\hat X_{\mu} E^{\Delta}_a \Big |_{E^{\Delta}_1=0,E^{\Delta}_2=0}=0.
\eea
The two equations (\ref{e12}) determine both the lattice and the difference equation. Both are constructed out of the invariants of the group $G$ prolonged to the lattice as indicated in eq. (\ref{e13}).  Thus, the difference scheme is by construction invariant under the entire group $G$ acting on the equation and lattice. In the continuous limit we have
\bea \label{e14}
E_1^{\Delta} = E + h_n E_1^{(1)} + h_{n+1} E_1^{(2)} +h_{n+2} E_1^{(3)} + \mbox{h.o.t.}, \\ \label{e15}
E_2^{\Delta} = 0 + h_n E_2^{(1)} + h_{n+1} E_2^{(2)} +h_{n+2} E_2^{(3)} + \mbox{h.o.t.}.
\eea
The terms spelled out in (\ref{e14}) correspond to the first differential approximation. 

Since the left hand side of (\ref{e14}, \ref{e15}) is invariant under $G$, the series on the right hand side must also be invariant. This does not guarantee that the first (or $n$-th) differential approximation will be invariant. In the next three sections we will show on examples that the first differential approximation is indeed invariant. Thus, choosing an invariant difference scheme guarantees, at least in the considered cases, that the aims of the differential approximation method are fully achieved.

The examples are all third order ODE's with 3 or 4 dimensional symmetry groups. In each case we write an invariant difference scheme of the form (\ref{e12}) and its first differential approximation (\ref{e14}). The terms $E^{(k)}_1$, ($k=1,2,3$) in (\ref{e14}) are differential expressions containing $y'''$ and $y''''$. These expressions will be simplified by removing $y'''$ and $y''''$, using the ODE (\ref{e4}) and its first differential consequence.
\section{Equations invariant under the similitude group $Sim(2)$.}
Let us consider the group of translations, rotations and uniform dilations of an Euclidean plane. Its Lie algebra $sim(2)$ is realized by the vector fields 
\bea \label{e15a}
\hat X_1 &=& \frac{\partial}{\partial x}, \quad 
\hat X_2 = \frac{\partial}{\partial y}, \quad
\hat X_3 = y \frac{\partial}{\partial x} - x\frac{\partial}{\partial y}, \\ \nonumber
\hat X_4 &=& x \frac{\partial}{\partial x} + y\frac{\partial}{\partial y}, 
\eea
This group has no second order differential invariant and precisely one third order one, namely\cite{26}
\begin{equation} \label{e16} 
I = \frac{(1+y^{'2}) y''' - 3y'y^{''2}}{y''^{2}}
\end{equation}
The expressions 
\begin{equation} \nonumber 
I_1 = \frac{y''}{(1 + y^{\prime 2})^{3/2}}, \quad
I_2 = \frac{(1+y^{'2}) y''' - 3y'y^{''2}}{(1 + y^{'2})^3}
\end{equation}
are invariant under the Euclidean group, with Lie algebra $\{ \hat X_1, \, \hat X_2, \, \hat X_3\}$, but only the ratio $I_2/I_1^2=I$ is invariant under dilations. 

Thus the lowest order ODE invariant under $Sim(2)$ is 
\begin{equation} \label{e17} 
(1 + y^{'2})y''' - 3y'y^{''2} = Ky^{''2}
\end{equation}
where $K$ is an arbitrary constant. 

To discretize (\ref{e17}) (or any third order ODE) we need (at least) a four-point stencil. The Euclidean group has 5 independent invariants depending on 4 points $(x_{n+k},y_{n+k}), k=-1,0,1,2$ namely \cite{26}
\begin{eqnarray}\label{e18}
\xi_1 
&= h_{n+2} \Bigg[ 1 + \bigg(\frac{y_{n+2}-y_{n+1}}{h_{n+2}}\Bigg)^2\Bigg]^{1/2},\nonumber\\
\xi_2 
&= h_{n+1} \Bigg[ 1 + \bigg(\frac{y_{n+1}-y_n}{h_{n+1}}\Bigg)^2\Bigg]^{1/2},\nonumber\\
\xi_3 
&= h_n \Bigg[ 1 + \bigg(\frac{y_n-y_{n-1}}{h_n}\Bigg)^2\Bigg]^{1/2},\\
\xi_4
&= (y_{n+2} - y_{n+1})h_{n+1} - (y_{n+1} - y_n)h_{n+2},\nonumber \\
\xi_5
&= (y_{n+1} - y_n)h_n - (y_n - y_{n-1})h_{n+1}. \nonumber 
\end{eqnarray}
Out of them we can construct 4 independent $Sim(2)$ invariants, for instance 
\begin{eqnarray}\label{e19}
J_1
&=& \frac{2\alpha \xi_4}{\xi_1\xi_2(\xi_1 + \xi_2)} + \frac{2\beta \xi_5}{(\xi_2 \xi_3)(\xi_2 + \xi_3)}\\ \nonumber
&&\alpha + \beta   = 1. \nonumber
\\ \nonumber
J_2 
&=& \frac{6}{\xi_1 + \xi_2 + \xi_3} \Biggl( \frac{\xi_4}{\xi_1\xi_2(\xi_1 + \xi_2)} - \frac{\xi_5}{\xi_2\xi_3(\xi_2 + \xi_3)} \Bigg) 
\end{eqnarray}
and the two ratios $\xi_1/\xi_2$, $\xi_2/\xi_3$. To obtain the continuous limit we put 
\bea \label{e20}
&&x_{n-1}=x_n-h_n, \; x_{n+1}=x_n + h_{n+1}, \\ \nonumber  &&x_{n+2}=x_n+h_{n+1}+h_{n+2}, \\ \nonumber
&&y_{n+k}=y(x_{n+k}), \; h_{n+k}=\alpha_k \epsilon, \, \alpha_k \sim 1,
\eea
and expand $y_{n+k}$ into a Taylor series about $x_n\equiv x$. 

The invariants $J_1$, $J_2$ where so chosen that their continuous limits are $I_1$ and $I_2$ respectively.

The invariant scheme used in \cite{26} to solve the ODE (\ref{e17}) numerically was
\bea \label{e21}
E^{\Delta}_2 &=& \xi_1 \xi_3 - \xi_2^2=0, \\ \label{e22}
E^{\Delta}_1&=&J_2 - K J_1^2=0.
\eea
The first differential approximation of (\ref{e21}) is
\bea \label{e23}
E^{\Delta}_2 &\approx& 2(-h_{n+1}^2 + h_n h_{n+2}) (y'^2+1)+ \\ \nonumber
&&+(2 h_n h_{n+1} h_{n+2} - 2 h_{n+1}^3 - h_n^2 h_{n+2} + \\ \nonumber && +h_n h_{n+2}^2) y' y''=0
\eea
Applying $\mbox{pr}^D \hat X_i$ to eq. (\ref{e23}) we find that the equation is invariant under the entire group $Sim(2)$, as are the terms of order $\epsilon^2$ and $\epsilon^3$ separately.

The first order differential approximation of the difference equation (\ref{e22}) is quite complicated. However, if we substitute the ODE (\ref{e17}) and its differential consequences into the first nonvanishing term of the approximation, we obtain a manageable expression 
\bea \label{e24}
E^{\Delta}_1 &\approx& (1+y'^2)y''' - 3 y' y''^2-Ky''^2-\\ \nonumber &-&\frac{1}{24} \frac{y''^3}{[1+y'^4](h_n+h_{n+1}+h_{n+2})} \times \\ \nonumber &\times& \Big \{ K^2 \big [ 16 \alpha (h_n+h_{n+1}+h_{n+2})^2 +
\\ \nonumber
&-&4 h_n^2 -12 h_{n+2}^2 - 8 h_{n+1}^2 -16 h_n h_{n+2}- \\ \nonumber &-&20 h_{n+1}h_{n+2} +12 h_nh_{n+1} \big ] + \\ \nonumber &+&9 h_{n+1}(h_{n+2}-h_n) \Big \} =0.
\eea
Expression (\ref{e24}) also satisfies
\beq \label{e25}
\mbox{pr}^{\Delta} \hat X_i E_1^{\Delta}  \Big |_{E_1^{\Delta} =E_2^{\Delta} =0}=0 \qquad i=1,\cdots,4
\eq
so the first differential approximation of the entire scheme (\ref{e21},\ref{e22}) is invariant under $Sim(2)$.

\section{Equations invariant under a one dimensional realization of $SL(2,\mathbb R)$.}

Four inequivalent realizations of $sl(2, \mathbb R)$ by vector fields in two variables exist \cite{29}. Invariant difference schemes for second and third order ODE's  have been constructed and tested for all of them \cite{26,27,28}. In this section we shall consider the first one, called $sl_1(2,\mathbb R)$, or alternatively $sl_y(2, \mathbb R)$, which actually involves one variable only:
\beq \label{e26}
\hat X_1 = \frac{\partial}{\partial y}, \quad \hat X_2 = y \frac{\partial}{\partial y}, \quad \hat X_3 = y^2 \frac{\partial}{\partial y}.
\eq
The corresponding Lie group acts by Mobius transformations (fractional linear transformations) on $y$.  The third order differential invariants of this action are the Schwarzian derivatives of $y$ and the independent variable x. The most general third order invariant ODE is 
\begin{equation} \label{e27}
\frac{1}{y^{'2}}\Biggl( y'y''' - \frac{3}{2} y^{''2}\Biggr) = F(x).
\end{equation}
where $F(x)$ is arbitrary. For $F(x)=K=const$ the group is $GL(2, \mathbb R)$ and (\ref{e26}) is extended by the vector field $\hat X_4 = \partial_x$. For $F(x)=0$ the symmetry group is $SL_y(2,\mathbb R) \otimes SL_x(2, \mathbb R)$.

The difference invariants on a four point stencil are 
\begin{equation} \label{e28}
R = \frac{(y_{n+2} - y_n)(y_{n+1} - y_{n-1})}{(y_{n+2} - y_{n+1})(y_n - y_{n-1})}, \quad x_n, h_{n+2}, h_{n+1}, h_n.
\end{equation}
The discrete invariant approximating the left hand side of (\ref{e27}) is 
\begin{eqnarray} \label{e29}
J_1
&= \frac{6h_{n+2} h_n}{h_{n+1}(h_{n+1}+h_{n+2})(h_n + h_{n+1})(h_{n+2} +h_{n+1} +h_n) } \\
&  \times \Biggl[ \frac{(h_{n+2} + h_{n+1})(h_{n+1} + h_n)}{h_n h_{n+2}} - R \Biggr] \nonumber
\end{eqnarray}
Any lattice depending only on $x_k$ will be invariant, in particular the lattice equation can be chosen to be
\beq \label{e30}
x_{n+1}-2x_n+x_{n-1}=0.
\eq
The general solution of (\ref{e30}) is
\beq \label{e31}
x_n = n h + c_0,
\eq
i.e. a uniform lattice with origin $x_0$ and spacing $x_{n+1}-x_n=h$.

Let us consider the invariant ODE
\begin{equation} \label{e32}
\frac{1}{y^{'2}}\Biggl( y'y''' - \frac{3}{2} y^{''2}\Biggr) = sin(x).
\end{equation}
and approximate it on an a priori arbitrary lattice. Such a scheme is given by 
\bea \label{e33}
&&E^{\Delta}=J_1-sin(\xi)=0,\\ \nonumber  &&\xi=x_n+a h_n+b h_{n+1} + c h_{n+2}, \\ \label{e34}
&&\phi(x_n, h_n, h_{n+1}, h_{n+2})=0,
\eea
where $a$, $b$, $c$ are constants and $\phi$ satisfies the condition
\beq \label{e35}
\phi(x_n, 0, 0, 0)\equiv 0,
\eq
(for instance $\phi$ can be linear as in (\ref{e30}).
The first differential approximation of (\ref{e33}), after the usual simplifications, is 
\bea \label{e36}
E_0 &\approx& \frac{1}{y'^2}\Biggl( y'y''' - \frac{3}{2} y^{''2}\Biggr)-sin(x) + \\ \nonumber &+& cos(x)\{ h_n(1+4a)-2*h_{n+1}(1-2b)-\\ \nonumber &-&h_{n+2}(1-4c)\}.
\eea
Eq. (\ref{e36}) is invariant under $SL_y(2,\mathbb R)$. Moreover, if we choose 
\beq \label{e37}
a=-\frac{1}{4},\; b=\frac{1}{2},\; c=\frac{1}{4},
\eq
the second term in (\ref{e36}) vanishes completely and $E_0=0$ is a second order approximation of the ODE (\ref{e32}). We mention that the uniform lattice (\ref{e31}), used in \cite{26} satisfies (\ref{e37}).
\section{Equations invariant under a two--dimensional realization of $GL(2,\mathbb R)$.}
A genuinely two-dimensional realization of the algebra $gl(2,\mathbb R)$ is given by the vector fields 
\bea\label{5.1}
\hat X_1 &=& \frac{\partial}{\partial y}, \quad 
\hat X_2 = x \frac{\partial}{\partial x} + y \frac{\partial}{\partial y},  \\ \nonumber 
\hat X_3 &=& 2xy \frac{\partial}{\partial x}+  y^2 \frac{\partial}{\partial y}, \quad
\hat X_4 = x\frac{\partial}{\partial x}.
\eea
The $SL(2, \mathbb{R})$ group corresponding to $\{\hat X_1, \hat X_2, \hat X_3\}$ has two differential invariants of order $m \le 3$:
\begin{equation}\label{5.2}
I_1 = \frac{2xy'' + y'}{y{'3}}, \quad I_2 = \frac{x^2(y' y''' - 3y^{''2})}{y^{'5}}.
\end{equation}
The expression $I_2/I_1^{\frac{3}{2}}$ is also invariant under the dilations generated by $\hat X_4$ and hence under $GL(2,\mathbb R)$. The corresponding ODE invariant under $GL(2,\mathbb R)$ is
\beq \label{5.3a}
I_2^2 = A^2 I_1^3
\eq
which is equivalent to
\begin{equation}\label{5.3}
E=x^4(y'y''' - 3y^{''2})^2 - A^2y' (2xy'' + y')^{3}=0.
\end{equation}

Five independent $SL(2,\mathbb R)$ difference invariants are
\begin{eqnarray} \label{5.4}
\xi_1 &=& \frac{1}{\sqrt{x_{n+1}x_{n+2}}} (y_{n+2} - y_{n+1}), \\ \nonumber 
\xi_2 &=& \frac{1}{\sqrt{x_n x_{n+1}}}(y_{n+1} - y_n),\\ \nonumber
\xi_3 &=& \frac{1}{\sqrt{x_{n-1}x_n}} (y_n - y_{n-1}), \\ \nonumber 
\xi_4 &=& \frac{1}{\sqrt{x_n x_{n+2}}}(y_{n+2} - y_n), \\
\xi_5 &=& \frac{1}{\sqrt{x_{n+1}x_{n-1}}} (y_{n+1} - y_{n-1}).  \nonumber
\end{eqnarray}
Any 4 ratios of $\xi_i/\xi_k$ will be $GL(2,\mathbb R)$ invariants. The $SL(2,\mathbb R)$ invariants that have the correct continuous limits are
\begin{eqnarray} \label{5.5}
J_2
&=& \frac{12}{\xi_2 (\xi_1 + \xi_2 + \xi_3)} \Big [ \frac{\xi_4 - \xi_1 - \xi_2}{\xi_1(\xi_1 + \xi_2)} -\\ \nonumber &-& \frac{\xi_5 -\xi_2 - \xi_3}{\xi_3(\xi_2 + \xi_3)} \Big ]\\
J_1 
&=& 8 \Biggl[ \alpha \frac{\xi_4 - \xi_1 - \xi_2}{\xi_1\xi_2(\xi_1 + \xi_2)} + (1-\alpha)  \frac{\xi_5 - \xi_2 - \xi_3}{\xi_2\xi_3(\xi_2 + \xi_3)}\Biggr] \nonumber\\
\end{eqnarray}
The difference scheme for the ODE (\ref{5.3}) with $A=-1$ used in Ref. \cite{26} was actually the square root of (\ref{5.3a}), equivalent to  
\bea \label{5.6}
E^{\Delta}_1 &=& J_2 + J_1^{\frac{3}{2}}=0, \\ \label{5.7}
E^{\Delta}_2 &=& \frac{\xi_1}{\xi_2}=\gamma = const.
\eea
The first differential approximation of the lattice equation (\ref{5.7}) is
\bea \label{5.8}
&&E^{(\Delta)}_1\approx (\gamma h_{n+1}-h_{n+2})y' +\\ \nonumber && \quad \frac{1}{2 x^2} \big \{ [- \gamma h_{n+1}^2
+ h_{n+2}(h_{n+2}+2 h_{n+1})]y' +\\ \nonumber &&\quad+ x y'' [\gamma h_{n+1}^2 - h_{n+1}h_{n+2} - h_{n+2}^2] \big \}=0.
\eea
Both terms in (\ref{5.8}) are invariant under $GL(2,\mathbb R)$.

The first differential approximation to the difference equation (\ref{5.6}) is
\bea \label{5.9}
E^{(\Delta)}_1&\approx& \frac{E}{y'^{10}} -\frac{\sqrt{y'}(y'+2xy'')^{\frac{7}{2}}}{16xy'^{10}(h_n+h_{n+1}+h_{n+2})}\times \\ \nonumber &\times&\Big [h_n^2(32\alpha-11)+16h_{n+1}^2(2\alpha-1)+\\ \nonumber &+&h_{n+2}^2(32\alpha-21)+h_nh_{n+1}(64\alpha-21)+\\ \nonumber &+&32h_nh_{n+2}(2\alpha-1)+h_{n+1}h_{n+2}(64\alpha-43) \Big ]
\eea
The second term of expression (\ref{5.9}) has been simplified using (\ref{5.3}) and its differential consequences. Again, (\ref{5.9}) is invariant under the entire $GL(2,\mathbb R)$ group.

\section{Conclusions}
The three examples considered above in Sections 3, 4 and 5 confirm that the method of invariant discretization provides a systematic way of constructing difference schemes for which the first differential approximation is invariant under the entire symmetry group of the original ODE.

The two equations (\ref{e12}) determining the invariant difference scheme for an ODE are not unique since there are more difference invariants than differential ones. The differential approximation of an invariant scheme can be used to benefit from this freedom and to choose an invariant scheme with a higher degree of accuracy. An example of this is given in Section 3, eq. (\ref{e36}) where the choice of the parameters  (\ref{e37}) assures that the terms of order $\epsilon$ in (\ref{e36}) vanish identically.

Previous numerical comparison between invariant discretization and standard noninvariant numerical methods for ODE's \cite{26,27,28} have shown two features. The first is that the discretization errors for invariant schemes are significantly smaller \cite{26} (by 3 orders of magnitude for eq. (\ref{e16}), 1 order of magnitude for eq. (\ref{5.3})). The second feature is that the qualitative behaviour of solutions close to singularities is described much more accurately by the invariant schemes \cite{26,27,28}.

An analysis of the relation between invariant discretization and the differential approximation method for PDE's is in progress.

\ack
We thank Alex Bihlo for interesting discussions. The research of P.W. was partially supported by a grant from NSERC of Canada. D.L. thanks the CRM for its hospitality. The research of D.L. has been partly supported by the Italian Ministry of Education and Research, 2010 PRIN "Teorie geometriche e analitiche dei sistemi Hamiltoniani in dimensioni finite e infinite". 

\section*{References}

\end{document}